\documentclass[submission,copyright,creativecommons]{eptcs}
\providecommand{\event}{QPL 2022}
\usepackage{breakurl}
\usepackage{amssymb,stmaryrd}
\usepackage{amsthm,amstext}
\usepackage{amsmath}
\usepackage{amsfonts}
\usepackage{MnSymbol,slashed}
\usepackage{fancyhdr}
\usepackage{graphicx}
\usepackage{pdfpages}
\usepackage{float}
\usepackage{subfig}
\usepackage{simplewick}
\usepackage{graphics,psfrag}
\usepackage{pstricks}
\usepackage{latexsym}
\usepackage{changepage}
\usepackage{caption}
\usepackage{paralist}
\usepackage{tikzit}

\tikzstyle{boundary vertex}=[inner sep=0mm, minimum size=1mm, shape=circle, draw=black, fill=black]
\tikzstyle{grey_dot}=[fill={rgb,255: red,191; green,191; blue,191}, draw={rgb,255: red,191; green,191; blue,191}, shape=circle, minimum size=1mm, inner sep=0mm]

\tikzstyle{arrow}=[->]
\tikzstyle{red_arrow}=[->, draw=red]
\tikzstyle{cyan_arrow}=[->, draw=cyan]
\tikzstyle{red_dash}=[-, dashed, draw=red]
\tikzstyle{grey dash}=[-, fill=none, draw={rgb,255: red,191; green,191; blue,191}, dashed]
\tikzstyle{dashed arrow}=[->, dashed]

\tikzstyle{gate}=[shape=rectangle, text height=1.5ex, text depth=0.25ex, yshift=0.5mm, fill=white, draw=black, minimum height=5mm, yshift=-0.5mm, minimum width=5mm, font={\small}, tikzit category=circuit]
\tikzstyle{big gate}=[shape=rectangle, text height=1.5ex, text depth=0.25ex, yshift=0.5mm, fill=white, draw=black, minimum height=10mm, yshift=-0.5mm, minimum width=5mm, font={\small}, tikzit category=circuit]
\tikzstyle{Z dot}=[inner sep=0mm, minimum size=2mm, shape=circle, draw=black, fill={rgb,255: red,221; green,255; blue,221}, tikzit category=zx]
\tikzstyle{Z phase dot}=[minimum size=5mm, font={\footnotesize\boldmath}, shape=rectangle, rounded corners=2mm, inner sep=0.2mm, outer sep=-2mm, scale=0.8, tikzit shape=circle, draw=black, fill={rgb,255: red,221; green,255; blue,221}, tikzit draw=blue, tikzit category=zx]
\tikzstyle{X dot}=[Z dot, shape=circle, draw=black, fill={rgb,255: red,255; green,136; blue,136}, tikzit category=zx]
\tikzstyle{X phase dot}=[Z phase dot, tikzit shape=circle, tikzit draw=blue, fill={rgb,255: red,255; green,136; blue,136}, font={\footnotesize\boldmath}, tikzit category=zx]
\tikzstyle{hadamard}=[fill=yellow, draw=black, shape=rectangle, inner sep=0.6mm, minimum height=1.5mm, minimum width=1.5mm, tikzit category=zx]
\tikzstyle{paulibox}=[fill={rgb,255: red,221; green,221; blue,255}, draw=black, shape=rectangle, inner sep=0.6mm, minimum height=5mm, minimum width=5mm, font={\footnotesize}, text height=1.5ex, text depth=0.25ex, tikzit category=zx]
\tikzstyle{vertex}=[inner sep=0mm, minimum size=1mm, shape=circle, draw=black, fill=black, tikzit category=misc]
\tikzstyle{vertex set}=[inner sep=0mm, minimum size=1mm, shape=circle, draw=black, fill=white, font={\footnotesize\boldmath}, tikzit category=misc]
\tikzstyle{small black dot}=[fill=black, draw=black, shape=circle, inner sep=0pt, minimum width=1.2mm, tikzit category=circuit]
\tikzstyle{cnot ctrl}=[fill=black, draw=black, shape=circle, inner sep=0pt, minimum width=1.2mm, tikzit category=circuit]
\tikzstyle{cnot targ}=[fill=white, draw=white, shape=circle, tikzit category=circuit, label={center:$\oplus$}, inner sep=0pt, minimum width=2.1mm, tikzit fill={rgb,255: red,102; green,204; blue,255}, tikzit draw=black]
\tikzstyle{ket}=[fill=white, draw=black, shape=regular polygon, regular polygon sides=3, regular polygon rotate=-30, scale=0.7, inner sep=1pt, tikzit category=circuit, tikzit shape=rectangle, tikzit fill=green]
\tikzstyle{bra}=[fill=white, draw=black, shape=regular polygon, regular polygon sides=3, regular polygon rotate=30, scale=0.7, inner sep=1pt, tikzit category=circuit, tikzit shape=rectangle, tikzit fill=red]
\tikzstyle{scalar}=[shape=rectangle, text height=1.5ex, text depth=0.25ex, yshift=0.5mm, fill=white, draw=black, minimum height=5mm, yshift=-0.5mm, minimum width=5mm, font={\small}]
\tikzstyle{clabel}=[fill=white, draw=none, shape=rectangle, tikzit fill={rgb,255: red,56; green,255; blue,242}, font={\footnotesize}, inner sep=1pt, tikzit category=labels]
\tikzstyle{empty diagram}=[draw={gray!40!white}, dashed, shape=rectangle, minimum width=1cm, minimum height=1cm, tikzit category=misc]

\tikzstyle{hadamard edge}=[-, dashed, dash pattern=on 2pt off 0.5pt, thick, draw={rgb,255: red,68; green,136; blue,255}]
\tikzstyle{box edge}=[-, dashed, dash pattern=on 2pt off 0.5pt, thick, draw={rgb,255: red,203; green,192; blue,225}]
\tikzstyle{brace edge}=[-, tikzit draw=blue, decorate, decoration={brace,amplitude=1mm,raise=-1mm}]
\tikzstyle{diredge}=[->]
\tikzstyle{double edge}=[-, double, shorten <=-1mm, shorten >=-1mm, double distance=2pt]
\tikzstyle{gray edge}=[-, {gray!60!white}]
\tikzstyle{pointer edge}=[->, very thick, gray]
\tikzstyle{boldedge}=[-, line width=1.6pt, shorten <=-0.17mm, shorten >=-0.17mm]
\usepackage{graphicx}
\usepackage{array}

\theoremstyle{plain}
\newtheorem{theorem}{Theorem}[section] 

\newtheorem{lemma}[theorem]{Lemma}

\newtheorem{definition}[theorem]{Definition}

\newtheorem{remark}[theorem]{Remark}
\newcommand{\CH}{\hbox{{$\mathcal H$}}}

\newcommand{\C}{\mathbb{C}}

\newcommand{\Z}{\mathbb{Z}}
\newcommand{\N}{\mathbb{N}}

\newcommand{\eps}{{\epsilon}}
\newcommand{\tens}{\mathop{{\otimes}}}
\newcommand{\la}{{\triangleright}}

\renewcommand{\>}{\rangle}

\def\lcross{{>\!\!\!\triangleleft}}

\newcommand{\vac}{{|{\rm vac}\>}}

\title{Qudit lattice surgery}
\author{Alexander Cowtan
\institute{Department of Computer Science,  University of Oxford\\
Wolfson Building, Parks Road, Oxford, UK}
\institute{Cambridge Quantum\\
Terrington House, 13-15 Hills Road, Cambridge, UK}
\email{akcowtan@gmail.com}}
\def\titlerunning{Qudit lattice surgery}
\def\authorrunning{Alexander Cowtan}

\begin{document}
\maketitle
\begin{abstract} 
We observe that lattice surgery, a model of fault-tolerant qubit computation, generalises straightforwardly to arbitrary finite-dimensional qudits. The generalised model is based on the group algebras $\mathbb{C}\mathbb{Z}_d$ for $d \geq 2$. It still requires magic state injection for universal quantum computation. We relate the model to the ZX-calculus, a diagrammatic language based on Hopf-Frobenius algebras.
\end{abstract}

\section{Introduction}
Topological quantum computing is a theoretical paradigm of large-scale quantum error correction in which important data is encoded in non-local features of a vast entangled state. So long as the physical errors on the overall system stay below some threshold value, the data is protected. The archetypal example is the $\C\Z_2$ surface code \cite{DKLP,BK1}, a system which requires only nearest-neighbour connectivity between qubits and has a high threshold against errors \cite{WFH}. The key practical feature of the surface code, as opposed to the earlier toric code \cite{Kit1}, is that it may be embedded on the plane with boundaries, and does not require exotic homology to encode data.

Lattice surgery was developed by Horsman et al. \cite{HFDM} as a method of computation using the surface code. It is conceptually simple and flexible, and believed to be efficient in its consumption of resources such as qubits and time \cite{FG,Lit1} compared to other methods such as defect braiding \cite{FMMC}. Lattice surgery starts with patches of surface code, then employs `splits' and `merges' on these, which act non-unitarily on logical states. In fact, merges are described by completely positive trace preserving (CPTP) maps, and cannot be performed deterministically in general. Both features make the model cumbersome to describe using the circuit model.

Interestingly, computation using lattice surgery closely mirrors the Hopf algebra structure of $\C \Z_2$, the group algebra of $\Z_2$ over the field $\C$. Coincidentally, this is one of the building blocks of the ZX-calculus, a formal graphical language for reasoning about quantum computation \cite{CD}. It represents quantum processes using ZX-diagrams, which may then be rewritten using the axioms of the calculus. The initial presentation of the ZX-calculus applied only to qubits, and can be summarised algebraically as $\C \Z_2$ and $\C(\Z_2)$ sitting on the same vector space, plus the inherent Fourier transform and a so-called phase group.

De Beaudrap and Horsman \cite{BH} noticed this relationship between lattice surgery and qubit ZX-calculus, and leveraged it to develop novel lattice surgery procedures. Other techniques using the same idea have also been developed, e.g. for efficient compilation of magic state distillation circuits \cite{GF} and reasoning about implementing deterministic programs despite non-deterministic merges \cite{BDHP}.

The ZX-calculus has since been generalised to qudits using $\C \Z_d$, for $d\in \N$ \cite{W1}. We observe that lattice surgery may similarly be generalised. The procedure is algebraically very simple, with the most advanced technology required being the Fourier transform. We give a concrete description of the computational model, although for brevity we elide some of the details such as the Pauli frame. We then leverage the qudit ZX-calculus to describe transformations on the logical data. We use this description to give a series of qudit lattice surgery procedures, and show that the model still requires magic state injection for universality.

\section{The $\C\Z_d$ surface code}
In this section we introduce the surface code for qudits; readers familiar with Kitaev models may wish to skip to Section~\ref{sec:lattice_surgery}. Throughout, we let $\Z_d$ be the cyclic group with $d$ elements labelled by integers $0,\cdots,d-1$ with addition as group multiplication. We assume $d\geq 2$, as the $d=1$ case is trivial. In the interest of brevity, proofs are brief or relegated to the appendices. For more explicit and thorough treatments at a higher level of generality see e.g. \cite{Kit1,Bom,Cow}. Throughout, we occasionally ignore normalisation (typically factors of $d$ or $1\over d$) when convenient.

\begin{definition} Let $\C\Z_d$ be the group Hopf algebra with basis states $|i\>$ for $i \in \Z_d$. $\C\Z_d$ has multiplication given by a linear extension of its native group multiplication, so $|i\>\otimes |j\> \mapsto |i+j\>$, and the unit $|0\>$. It has comultiplication given by $|i\>\mapsto |i\>\otimes |i\>$, and the counit $|i\>\mapsto 1 \in \C$. It has the normalised integral element $\Lambda_{\C\Z_d} = \frac{1}{d}\sum_i |i\>$ and the antipode is the group inverse. $\C\Z_d$ is commutative and cocommutative.
\end{definition}

\begin{definition}Let $\C(\Z_d)$ be the function Hopf algebra with basis states $|\delta_i\>$ for $i\in \Z_d$. $\C(\Z_d)$ is the dual algebra to $\C\Z_d$. $\C(\Z_d)$ has multiplication $|\delta_i\> \otimes |\delta_j\> \mapsto \delta_{i,j}|\delta_i\>$ and the unit $\sum_i |\delta_i\>$. It has comultiplication $|\delta_i\> \mapsto \sum_{h\in \Z_d} |\delta_h\>\otimes|\delta_{i-h}\>$ and counit $|\delta_i\>\mapsto \delta_{i,0}$. It has the normalised integral element $\Lambda_{\C(\Z_d)} = |\delta_0\>$ and the antipode is also the inverse. $\C(\Z_d)$ is commutative and cocommutative.
\end{definition}

\begin{lemma}\label{lem:fourier}
The algebras are related by the Fourier isomorphism, so $\C(\Z_d)\cong \C\Z_d$ as Hopf algebras. In particular this isomorphism has maps
\begin{equation}\label{Zisom} |j\> \mapsto \sum_k q^{jk}|\delta_k\>,\quad |\delta_j\>\mapsto {1\over d} \sum_k q^{-jk}|k\>,\end{equation}
where $q = e^{i2\pi\over d}$ is a primitive $d$th root of unity.
\end{lemma}

\begin{definition}\label{def:lattice_acts}
Now let $\Sigma = \Sigma(V, E, P)$ be a square lattice viewed as a directed graph with its usual (cartesian) orientation. The corresponding Hilbert space $\CH$ will be a tensor product of vector spaces with one copy of $\C\Z_d$ at each arrow in $E$, with basis denoted by $\{|i\>\}_{i\in \Z_d}$ as before. Next, for each vertex $v \in V$ and each face $p \in P$ we define an action of $\C\Z_d$ and $\C(\Z_d)$, which acts on the vector spaces around the vertex or around the face, and trivially elsewhere, according to
\[\input{vertex_action.tikz}\]
and
\[\input{face_action.tikz}\]
for $|l\> \in \C\Z_d$ and $|\delta_j\>\in \C(\Z_d)$.
\end{definition} 

Here $|l\>\la_v$ subtracts in the case of arrows pointing towards the vertex and $|\delta_j\>\la_p$ has $c,d$ entering negatively in the $\delta$-function because these are contra to a {\em clockwise} flow around the face in our conventions. The vertex actions are built from four-fold copies of the operator $X$ and $X^\dagger$, where $X^l|i\>=|i+l\>$. Consider the face actions of elements $\sum_j q^{mj}|\delta_j\>$, i.e. the Fourier transformed basis of $\C(\Z_d)$; these face actions are made up of $Z$ and $Z^\dagger$, where $Z^m|i\>=q^{mi}|i\>$, and the $Z$, $X$ obey $ZX=qXZ$.

Stabilisers on the lattice are given by measurements of the $X\otimes X\otimes X^\dagger\otimes X^\dagger$ and $Z\otimes Z\otimes Z^\dagger\otimes Z^\dagger$ operators on vertices and faces respectively; that is, for the vertices we non-deterministically perform one of the $d$ projectors $P_v(j) = \sum_k q^{jk}|k\>\la_v$ for $j\in \Z_d$, according to each of the $d$ measurement outcomes. Similarly for faces, we perform one of the $d$ projectors $P_p(j) =|\delta_j\>\la_p$. In practice, this requires additional `syndrome' qudits at each vertex and face; we give explicit circuits for these in Appendix~\ref{app:circs}. At each round of measurement, we measure all of the stabilisers on the whole lattice. Physically, we may also say that the system is in a subspace of a certain Hamiltonian:
\[H=-(\sum_v A(v) + \sum_p B(p))+{\rm const.}\]
where
\[ A(v)=P_v(0)=\Lambda\la_v={1\over d}\sum_i |i\>\la_v,\quad B(p)=P_p(0)=\Lambda^*\la_p=|\delta_0\>\la_p.\]
It is easy to see that 
\[ A(v)^2=A(v),\quad B(p)^2=B(p),\quad [A(v),A(v')]=[B(p),B(p')]=[A(v),B(p)]=0\]
where $v,v'$ are different vertices, and $p,p'$ are different faces.

When the measurements at a vertex $v$ and face $p$ yield the projectors $A(v)$ and $B(p)$ we say that no errors were detected at these locations, and we are locally in the vacuum. Then if we obtain the projectors $A(v)$ and $B(p)$ everywhere we are in the global vacuum space $\CH_{vac}$. One can check that a state $\vac \in \CH_{vac}$ obeys
\[|l\>\la_v\vac = A(v)\vac= \sum_j q^{mj}|\delta_j\>\la_p\vac = B(p)\vac=\vac\]
for all $l, m\in \Z_d, v\in V, p\in P$.
\begin{definition}\label{def:log_states}
We can always write down at least two vacuum states\footnote{In certain cases, such as when the lattice is embedded onto a sphere, these states coincide.}, which we shall call:
\[|0\>_L := \prod_{v}A(v)\bigotimes_E |0\>\]
and
\[|\delta_0{}\>_L := \prod_{p}B(p)\bigotimes_E \sum_i |i\>.\]
\end{definition}
Computationally, the vacuum space is also the \textit{logical space}, the subspace in which we store data; the subscript ${}_L$ refers to this logical space, and $|0\>_L$, $|\delta_0{}\>_L$ are canonical logical states. 

If measurements yield other projectors $P_v(j)$ or $P_p(j)$ then we have detected an error; in physics jargon we have detected the presence of an electric or magnetic particle. One important feature of the code is that if we receive the measurement outcome $P_v(j)$, say, at a vertex then there will be another vertex at which we detect $P_v(-j)$ instead. This is because all operators on the lattice come in the form of \textit{string operators}. String operators come in two types: $X$ and $Z$.

\begin{definition} 
An $X$-type string operator ${}_xF^i_\xi$ acts on the lattice as
\[\input{x_string_operator.tikz}\]
where $\xi$ is a string that passes between faces and for each crossed edge we apply either an $X^i$ or $X^i{}^\dagger$ depending on the orientation, as shown. 
\end{definition}
The $X$-type string operators satisfy $({}_xF^i_{\xi})^\dagger = {}_xF^{-i}_{\xi}$. Additionally we have
\[{}_xF^i_{\xi}\circ{}_xF^j_{\xi} = {}_xF^{i+j}_{\xi}\]
and, given concatenated strings $\xi, \xi'$,
\[{}_xF^i_{\xi'\circ\xi} = {}_xF^i_{\xi'}\circ{}_xF^i_{\xi}\]
where one can see multiplication and comultiplication of $\C \Z_d$; more generally they obey the same Hopf laws as $\C \Z_d$. The other axioms are easy to check.

The $X$-type string operators make magnetic quasiparticles `appear' at the faces at which a string ends. In particular, given an initial vacuum state $\vac$, we can check that 
\[P_{p_0}(i){}_xF^i_\xi\vac = P_{p_1}(-i){}_xF^i_\xi\vac = {}_xF^i_\xi\vac\]
where $p_0,p_1$ are the start and endpoints of the string, so we will detect errors at these locations upon measurement. However, the string operators leave the system in the vacuum in the intermediate faces of the string, as we have:
\[B(p){}_xF^i_\xi = {}_xF^i_\xi B(p)\]
for any $p \neq p_0$ or $p_1$. As a consequence, we may think of string operators as equivalent up to a sort of discrete framed isotopy.

\begin{definition}
A $Z$-type string operator ${}_zF^{\delta_j}_\xi$ acts on the lattice as
\[\input{z_string_operator.tikz}\]
by passing between vertices. For each crossed edge we include a term in the $\delta$-function, as shown. Observe that $\sum_j q^{mj} {}_zF^{\delta_j}_\xi$ applies a $Z^m$ or $Z^m{}^\dagger$ at each edge, and that this is the Fourier transformed basis of the $Z$-type string operators. 
\end{definition}
The $Z$-type string operators satisfy $({}_zF^{\delta_i}_\xi)^\dagger = {}_zF^{\delta_i}_\xi$. Additionally we have
\[{}_zF^{\delta_i}_\xi\circ {}_zF^{\delta_j}_\xi = \delta_{i,j}\ {}_zF^{\delta_j}_\xi\]
and
\[{}_zF^{\delta_i}_{\xi'\circ\xi} = \sum_{_h}{}_zF^{\delta_h}_{\xi'}\circ{}_zF^{\delta_{i-h}}_{\xi},\]
so $Z$-type string operators obey the same Hopf laws as $\C(\Z_d)$.

The $Z$-type string operators generate electric quasiparticles at the vertices at which a string ends. We have
\[P_{v_0}(i)\sum_j q^{ij}{}_zF^{\delta_j}_\xi\vac=P_{v_1}(-i)\sum_j q^{ij}{}_zF^{\delta_j}_\xi\vac=\sum_j q^{ij}{}_zF^{\delta_j}_\xi\vac.\]
As a result, we refer to this basis of the $Z$-type string operators as the `quasiparticle basis'. They leave the system in the vacuum in the intermediate vertices of the string:
\[A(v)\sum_j q^{ij}{}_zF^{\delta_j}_\xi = \sum_j q^{ij}{}_zF^{\delta_j}_\xi A(v)\]
for any $v\neq v_0$ or $v_1$.

In the quasiparticle basis we have
\[\sum_j q^{ij}{}_zF^{\delta_j}_\xi\circ\sum_j q^{kj}{}_zF^{\delta_j}_\xi = \sum_j q^{(i+k)j}{}_zF^{\delta_j}_\xi\]
and
\[\sum_j q^{ij}{}_zF^{\delta_j}_{\xi'\circ\xi} = \sum_j q^{ij}{}_zF^{\delta_j}_{\xi'}\circ \sum_k q^{ik}{}_zF^{\delta_k}_{\xi}\]
i.e. the same algebraic rules as the $X$-type string operators, and as $\C \Z_d$.

\begin{lemma}\cite{Kit1}
String operators which form a closed loop on a lattice segment which is locally vacuum either act as identity or are physically impossible, i.e. they take the system to 0.
\end{lemma}
\proof
First, assume the string passes between faces, so we have an $X$-type string operator. In this case, we may tile the loop with squares on the dual lattice (that is, the dual in the graph-theoretic sense). Then one can check the closed string operator acts as a product over the tiles of $|l\>\la_v$ actions. As $|l\>\la_v\vac = \vac$ the state is left unchanged.

If the string passes between vertices, tile the loop with squares. Consider the $Z$-type string operators in the quasiparticle basis. Then the closed string operator acts as a product of $\sum_j q^{mj}|\delta_j\>\la_p$ actions. As $\sum_j q^{mj}|\delta_j\>\la_p\vac = \vac$ the state is left unchanged. In the original basis, the product of $|\delta_h\>\la_p$ actions acts as identity if $h=0$; otherwise it takes the system to 0.
\endproof

We are now ready to define a \textit{patch}.

\begin{definition}
A patch is a rectangular segment of lattice bordered by two rough and two smooth boundaries, like so:
\[\input{patch.tikz}\]
where rough boundaries are at the top and bottom, while smooth boundaries are at the left and right.\footnote{One can of course define patches with other combinations of boundaries, which are useful for specific kinds of circuits \cite{Lit1}, but this is a convenient definition for our purposes.}
\end{definition}

There are assumed to be no parts of the lattice beyond the patch; these are all of the edges in the lattice. The stabilisers on the boundaries are the same as in the bulk, with the exceptions that (a) stabilisers obviously exclude the edges which are not present, and (b) there are no stabilisers for single edges. So, in particular, there are no vertex measurements which include only the single top and bottom edges; likewise, there are no face measurements which include only the single left and right edges.\footnote{More generally, boundary conditions are defined by a subgroup $K\subseteq\Z_d$. This leads to a rich algebraic theory \cite{BSW}. In the present case, the subgroups $K$ associated to rough and smooth boundaries are $K=\{0\}$ and $K=\Z_d$ respectively.}

\begin{lemma}
Let the system be in a vacuum state. All $X$-type string operators which extend between the left to right boundaries, for example in the manner below
\[\input{patch_X_string.tikz}\]
leave the system in vacuum, but do not generally act as identity.
\end{lemma}
\proof
There are no face stabilisers for the single edges at the end, and at all other faces $B(p)$ commutes with the string operators. However, while the string can be smoothly deformed up and down the sides of the patch while leaving the operation on the vacuum invariant, it cannot be expressed as a product of vertex or face operators, and explicit checks on small (but nontrivial) examples show that ${}_xF^i_{\xi}$ does not act as identity unless $i=0$.
\endproof

In fact, we have a stronger property: all operators which act as a product of $X$ operations on edges and leave the system in vacuum may be expressed as a linear combination of $X$-type string operators extending between left and right, so the $d$ different $X$-type string operators ${}_xF^i_{\xi}$ form an orthonormal basis for the algebra of such operators. We have a similar result for the $Z$-type string operators in the quasiparticle basis, which extend between the top and bottom boundaries. These properties motivate the following:
\begin{lemma}
A patch as defined above with underlying group algebra $\C \Z_d$ has ${\rm dim}(\CH_{vac}) = d$ and two canonical bases, $\{|i\>_L\}_{i\in\Z_d}$ and $\{|\delta_i\>_L\}_{i\in \Z_d}$.
\end{lemma}
\proof
The states in $\CH_{vac}$, and hence the logical space of the code, are uniquely characterised by the algebra of operators upon them. Given a reference state $|{\rm ref}\>$ in the vacuum, if there is another vacuum state $|\psi\>$ there must be some linear map which transforms $|{\rm ref}\>$ into $|\psi\>$. Thus $\{{}_xF^i_{\xi}\}_{i\in\Z_d}$ automatically gives an orthonormal basis for $\CH_{vac}$.

Let us call $|0\>_L$ the reference state from Def~\ref{def:log_states}. Then $|i\>_L := {}_xF^i_{\xi}|0\>_L$, where $\xi$ is any string extending from the left boundary to the right. We may call ${}_xF^i_{\xi}$ a logical $X^i$ gate, i.e. $X^i_L$.

As with $\C\Z_d$ itself, we have a Fourier basis for the patch's logical space. To begin with, we have $|\delta_0\>_L = \sum_i {}_xF^i_{\xi}|0\>_L$. Then, we define further logical states in the Fourier basis by $|\delta_i\>_L = \sum_j q^{ij}{}_zF^{\delta_j}_{\xi'}|\delta_0\>_L$, where now the string $\xi'$ extends from the top to bottom, and we claim that $|\delta_i\>_L = \sum_kq^{-ik}|k\>_L$. We check on a small example that this is consistent with Lemma~\ref{lem:fourier} in Appendix~\ref{app:fourier_patch}, and assert that it holds generally. As a result we call $\sum_j q^{ij}{}_zF^{\delta_j}_{\xi'}$ a logical $Z^i$ gate, $Z^i_L$.
\endproof
Note that the logical space is independent of the size of lattice, and depends only on the topology. The lattice size is relevant only for the probability of correcting errors.

\section{Lattice surgery}\label{sec:lattice_surgery}
If we have two patches with logical spaces $(\CH_{vac})_1$ and $(\CH_{vac})_2$ which are disjoint in space then we evidently have a combined logical space $\CH_{vac} = (\CH_{vac})_1 \tens (\CH_{vac})_2$.

We may start with one patch and `split' it to convert it into two patches.
\subsection{Splits}
To perform a smooth split, take a patch and measure out a string of intermediate qudits from top to bottom in the $\{|\delta_i\>\}$ basis, like so:
\[\input{split1.tikz}\]
Regardless of the measurement results we get, we now have two disjoint patches next to each other. We can see the effect on the logical state by considering an $X$-type string operator which had been extending across a string $\xi$ from left to right on the original patch. Previously it had been ${}_xF^i_{\xi}$, say. Now, let $\xi = \xi''\circ\xi'$, where $\xi'$ extends across the left patch after the split and $\xi''$ extends across right one. Then ${}_xF^i_{\xi} = {}_xF^i_{\xi'}\circ{}_xF^i_{\xi''}$; our $X^i_L$ gate on the original logical space is taken to $X^i_L\tens X^i_L$ on $(\CH_{vac})_1 \tens (\CH_{vac})_2$. It is easy to see that this then gives the map:
\[\Delta_s : |i\>_L\mapsto |i\>_L\otimes |i\>_L\]
for $i\in\Z_d$. This is the same regardless of the measurement outcomes on the intermediate qubits we measured out.

To perform a rough split, take a patch and measure out a string of qudits from left to right in the $\{|i\>\}$ basis. A similar analysis to before, but for $Z^i_L$ gates, shows that we have
\[\Delta_r : |\delta_i\>_L\mapsto |\delta_i\>_L\otimes |\delta_i\>_L.\]
\begin{remark}\rm
We now note a subtlety: for both smooth and rough splits we induce a copy in the relevant bases, that is the comultiplication of $\C\Z_d$, rather than the comultiplication of $\C(\Z_d)$ for the rough splits. This is because we are placing both algebras on the same object, using the non-natural isomorphism $V\cong V^*$ for vector spaces $V$. Thus if we take the rough split map in the other basis we get
\[\Delta_r : |i\>_L\mapsto \sum_h |h\>_L \otimes |i-h\>_L.\]
This follows directly from Lemma~\ref{lem:fourier}. The fact that both algebras are placed on the same object allows us to relate the model to the ZX-calculus in Section~\ref{sec:zx}.
\end{remark}

\subsection{Merges}
To perform a smooth merge, we do the reverse operation. Start with two disjoint patches:
\[\input{split2.tikz}\]
and then initialise between them a string of intermediate qudits, each in the $\sum_i|i\>$ state, like so:
\[\input{merge.tikz}\]
Then measure the stabilisers at all sites on the now merged lattice. Now, assuming no errors have occurred all the stabilisers are automatically satisfied everywhere except the measurements which include the new edges. These measurements realise a measurement of $Z_L\tens Z_L$ on the logical space $(\CH_{vac})_1 \tens (\CH_{vac})_2$. We prove this in Appendix~\ref{app:merge}. With merges, the resultant logical state after merging is also dependent on the measurement outcomes. 

Depending on which `frame' we choose we can have two different sets of possible maps from the smooth merge, see \cite{BH} for the easier qubit case. Here we choose to adopt the Pauli frame of the second patch. In the Fourier basis we thus have the Kraus operators:
\[\nabla_s: \{|\delta_i\>_L\tens|\delta_j\>_L\mapsto q^{in}|\delta_{i+j}\>_L\}_{n \in \{0,\cdots,d-1\}}\]
where $q^{in}$ is a factor introduced by the $Z_L\tens Z_L$ measurement; we have $n \in \{0,\cdots,d-1\}$ for the $d$ different possible measurement outcomes. If we only consider the $n=0$ case for a moment, one can come to the conclusion that this is the correct map using the $Z_L$ logical operators:
\[\sum_j q^{ij}{}_zF^{\delta_j}_\xi\circ\sum_j q^{kj}{}_zF^{\delta_j}_\xi = \sum_j q^{(i+k)j}{}_zF^{\delta_j}_\xi\]
from earlier, where $\xi$ extends from bottom to top on both original patches. Then when we merge the patches, we get the combined string operator. In the other basis of logical states, the smooth merge gives:
\[\nabla_s: \{|i\>_L\tens|j\>_L\mapsto \delta_{i+n,j}|i+n\>_L\}_{n\in \{0,\cdots,d-1\}},\]

\begin{remark}\rm
It is common in categorical quantum mechanics to consider the so-called multiplicative fragment of quantum mechanics. In this fragment, we may post-select rather than just make measurements according to the traditional postulates. As such, there is a choice of post-selection such that $n=0$ and we acquire the multiplication of $\C\Z_d$ or $\C(\Z_d)$ depending on basis. While physically we cannot post-select, this is a useful toy model in which algebraic notions may be more conveniently related to quantum mechanical processes.
\end{remark}

Considering the same convention of frame, a rough merge gives:
\[\nabla_r: \{|i\>_L\tens|j\>_L\mapsto q^{in}|i+j\>\}_{n\in \{0,\cdots,d-1\}}\]
by a similar argument, this time performing a measurement of $X_L\tens X_L$ to merge patches at the top and bottom.
\subsection{Units and deletion}
While we are on the subject of measurements, we can delete a patch by measuring out every qudit associated to its lattice in the $Z$-basis. If we do so, we obtain the maps
\[\eps_r: \{|i\>_L\mapsto \delta_{n,i}\}_{n\in \{0,\cdots,d-1\}}.\]
In the $n=0$ outcome this is precisely the counit of $\C(\Z_d)$. We check this in Appendix~\ref{app:counit}. If we instead measure out each qudit in the $X$-basis we get
\[\eps_s: \{|i\>_L\mapsto q^{in}\}_{n\in \{0,\cdots,d-1\}},\]
where we see the counit of $\C\Z_d$.

One can clearly also construct the units of $\C(\Z_d)$ and $\C\Z_d$, being $\eta_s: \sum_i|i\>_L$ and $\eta_r: |0\>_L$ respectively. The last remaining pieces of the puzzle are the antipode and Fourier transform on the logical space.
\subsection{Antipode}
First we demonstrate how to map between the $|0\>_L$ and $|\delta_i\>_L$ states. If we apply a Fourier transform $H = \sum_{j,k}q^{-jk}|k\>\<j|$ to a qudit in the state $|0\>$ we have $H|0\> = \sum_i |i\>$.\footnote{The $H$ stands for Hadamard, which is what the qubit Fourier transform is commonly called. The qudit Fourier transform is not a Hadamard matrix in general.} As $HX = Z^\dagger H$ (and $XH = HZ$) all $A(v)$ projectors are translated to $B(p)$ projectors by rotating the lattice to exchange vertices with faces
\[\input{vertex_rotate.tikz}\]
such that the $X, X^\dagger$ match up with $Z^\dagger, Z$ appropriately when considering the clockwise conventions from Def~\ref{def:lattice_acts}.
This is just a conceptual rotation, and there does not need to be any \textit{physical} rotation in space. Thus we have
\[H_L |0\>_L=(\bigotimes_E H) \prod_{v}A(v)\bigotimes_E |0\> = \prod_pB(p)\bigotimes_E \sum_i |i\> = |\delta_i\>_L\]
where $H_L = \bigotimes_E H$ is the logical Fourier transform, and the lattice has been mapped:
\[\input{rotate_patch.tikz}\]

$H_L$ also takes $X$-type string operators to $Z$-type string operators in the quasiparticle basis but with a sign change, and thus we have
\[H_L|i\>_L = H_LX^i|0\>_L=Z^{-i}H_L|0\>_L = \sum_{k}q^{-ik}|k\>_L=|\delta_i\>_L\]
so it is genuinely a Fourier transform. Applying it twice gives
\[H_LH_L|i\>_L = \sum_{k,l}q^{-ik}q^{-kl}|l\>_L = \sum_l \delta_{l,-i}|l\>_L = |-i\>_L\]
where the lattice is now as though the whole patch has been rotated in space by $\pi$ by the same argument as before. This is evidently the \textit{logical antipode}, $S_L = H_LH_L$.

This completes the set of fault-tolerant operations we may perform with the $\C\Z_d$ lattice surgery. One can create other states in a non-error corrected manner and then perform state distillation to acquire the correct state with a high probability, but this is beyond the scope of the paper and very similar to e.g. \cite{FSG}.

\section{The ZX-calculus}\label{sec:zx}
The ZX-calculus is based on Hopf-Frobenius algebras sitting on the same object. It imports ideas from monoidal category theory to justify its graphical formalism \cite{Sel}. See \cite{HV} for an introduction from the categorical point of view. Calculations may be performed by transforming diagrams into one another, and the calculus may be thought of as a tensor network theory equipped with rewriting rules.

Here we present the syntax and semantics of ZX-diagrams for $\C\Z_d$. We are unconcerned with either universality or completeness \cite{Back}, and give only the necessary generators for our purposes; moreover, we adopt a slightly simplified convention. First, we have generators:
\[\input{units.tikz}\]
for elements, where the small red and green nodes are called `spiders', and diagrams flow from bottom to top.\footnote{Red and green are dark and light shades in greyscale.} The labels associated to a spider are called phases. Then we have the multiplication maps,
\[\input{merge_spiders.tikz}\]
comultiplication,
\[\input{comult_spiders.tikz}\]
maps to $\C$,
\[\input{counits.tikz}\]
and Fourier transform\footnote{The Hadamard symbol here makes it look like it is vertically reversible, i.e. $H^\dagger = H$, but it is not; this is just a notational flaw.} plus antipode:
\[\input{hadamard.tikz}\]
Now, these generators obey all the normal Hopf rules: associativity of multiplication and comultiplication, unit and counit, bialgebra and antipode laws, but that it is not all. The ZX-calculus makes use of an old result by Pareigis \cite{Par}, which states that all finite-dimensional Hopf algebras on vector spaces automatically give two Frobenius structures, which in the present case correspond to the red and green spiders above. In this case, they are in fact so-called $\dagger$-special commutative Frobenius algebras ($\dagger$-SFCAs) \cite{CPV}. Such algebras have a normal form, such that any connected set of green or red spiders may be combined into a single green or red spider respectively, summing the phases \cite{CD}. This is called the \textit{spider theorem}. As an easy example, observe that we can define the $X^a$ gate in the ZX-calculus as:
\[\input{X_gate_spider.tikz}\]
and similarly for a $Z^b$ gate,
\[\input{Z_gate_spider.tikz}.\]
The Fourier transform then `changes colour' between green and red spiders. We show these axioms in Appendix~\ref{app:zx_axioms}. For a detailed exposition of the qudit ZX-calculus in greater generality see \cite{W1}. 

Now, one can immediately see that the generators are automatically (by virtue of the $\C\Z_d$ and $\C(\Z_d)$ structures) in bijection with the lattice surgery operations described previously. The bijection between this fragment of the ZX-calculus and lattice surgery was spotted by de Beaudrap and Horsman in the qubit case \cite{BH}; however, their presentation emphasises the Frobenius structures. The algebraic explanation for the lattice surgery properties is all in the Hopf structure: in summary, it is because the string operators are Hopf-like.\footnote{We formalise such operators as module maps in \cite{Cow}.} The Frobenius structures are still useful diagrammatic reasoning tools because of the spider theorem, and also because the two interacting Frobenius algebras correspond to the rough (red spider) and smooth (green spider) operations. There is a convenient 3-dimensional visualisation for this using `logical blocks', which we defer to Appendix~\ref{app:block}. There we also include Table~\ref{tbl:lat_oper}, which is a dictionary between lattice operations, ZX-diagrams and linear maps.

\subsection{Gate synthesis}\label{sec:synth}
Using the ZX-calculus we can thus design logical protocols in a straightforward manner. We have already implicitly shown a  state injection protocol, being the spider merges for the $X^a$ and $Z^b$ gates above, but we can go further. A common gate in the circuit model is the controlled-$X$ ($CX$) gate. In qudit quantum computing this is defined as the map 
\[CX: |i\> \otimes |j\> \mapsto |i\> \otimes |i+j\>\]
which in the ZX-calculus we might represent as, say,
\[\input{cnot_spiders.tikz}.\]
In the first diagram we perform a smooth split followed by a rough merge; in the second we do the opposite. In the third and fourth we first generate a maximally entangled state and then perform a smooth and rough merge on either side. The antipodes are necessary because of a minor complication with duals in the qudit ZX-calculus. Rewrites using the calculus show that these are equal, and conversions into linear maps do indeed yield the $CX$. We check this in Appendix~\ref{app:cx}. Note that we implicitly assumed the $n=0$ measurement outcomes for the merges, but we assert that in this case the protocol works deterministically by applying corrections. This is a generalisation of protocols specified in \cite{BH}, and the correction arguments are identical.

We can also easily see that the lattice surgery operations are not universal, even with the addition of logical $X_L$ and $Z_L$ gates using string operators. All phases have integer values and so we cannot even achieve all single-qudit gates in the 2nd level of the Clifford hierarchy fault-tolerantly. For example, we cannot construct a $\sqrt{X}_L$ gate with the operations listed here.

With this limitation in mind, in Appendix~\ref{app:generalisations} we discuss the prospects for expanding the scope of the model to other group algebras and to Hopf algebras more generally.

\section{Conclusion}
We have shown that lattice surgery is straightforward to generalise to qudits, assuming an underlying abelian group structure. The resultant diagrammatics which can be used to describe computation are elegant, concise and powerful. We currently do not know how this generalises further, and what the connections are to quantum field theories. We aim to tackle these issues in  future work. 

\section{Acknowledgements}
We thank the Wolfson Harrison UK Research Council Quantum Foundation Scholarship for making this work possible.

\appendix
\section{Circuits for measuring stabilisers}\label{app:circs}
Given the face
\[\input{face_to_be_measured.tikz}\]
we can perform a face measurement using the circuit
\[\input{Z_stab_measurement.tikz}\]
i.e. a measurement of the $Z\tens Z\tens Z^\dagger\tens Z^\dagger$ operator. The $CX$ gates act as $|i\>\tens |j\>\mapsto |i\>\tens|i+j\>$, and the yellow boxes are Fourier transforms $H$. Note that $H^2 : |i\>\mapsto |-i\>$. Hence one can calculate that this circuit is the map
\[|i\>\tens|j\>\tens|k\>\tens|l\>\mapsto \delta_a(i+j-k-l)|i\>\tens|j\>\tens|k\>\tens|l\> \]
for some $a\in \Z_d$. For a vertex
\[\input{vertex_to_be_measured.tikz}\]
we have
\[\input{X_stab_measurement.tikz}\]
measuring the $X\tens X\tens X^\dagger\tens X^\dagger$ operator. The $CX$ gates also act as $|\delta_i\>\tens|\delta_j\>\mapsto |\delta_{i-j}\>\tens|\delta_j\>$, motivating the exchanged control and target and application of $H^2$ to the other qudits. Then we see that this circuit is the map
\[|\delta_i\>\tens|\delta_j\>\tens|\delta_k\>\tens|\delta_l\>\mapsto \delta_b(i+j-k-l)|\delta_i\>\tens|\delta_j\>\tens|\delta_k\>\tens|\delta_l\>\]
for some $b\in Z_d$.

\section{Fourier basis for patches}\label{app:fourier_patch}
Consider the small patch
\[\input{small_patch.tikz}\]
Now, $|i\>_L$ is the following state:
\[\input{patch_0.tikz}\]
where we have taken $|0\>_L$ and applied an $X$-type string from left to right. Now, consider $|\delta_0\>_L$:
\[\input{patch_plus.tikz}\]
where we performed a change of variables $g\mapsto -g$, $h\mapsto -h$. Now, $\delta_0(d+a-c-f-g+h)$ holds iff $d+a-g=i$ and $-f-c+h=-i$ for some $i\in \Z_d$. Thus we have $|\delta_0\>_L = \sum_i|i\>_L$. If we then apply a $Z$-type string operator from top to bottom in the quasiparticle basis we see that $|\delta_j\>_L = \sum_iq^{-ij}|i\>_L$.

One could then show that the bases are consistent under Fourier transform for all sizes of patch by induction, using the above as the base case.

\section{Proof of lattice merges}\label{app:merge}
We demonstrate the smooth merge on a small patch but it is easy to see that the same method applies for arbitrary large patches. We begin with two patches, in the $|\delta_g\>_L$ and $|\delta_h\>_L$ states respectively.
\[\input{patch_delta_0.tikz}\]
Then initialise two new edges between, each in the $|\delta_0\>$ state.
\[\input{patch_delta_0_together.tikz}\]
where we have exaggerated the length of the new edges for emphasis. Now if we apply stabiliser measurements at all points we see that the only relevant ones are the face measurements including the new edges (the vertex measurements will still yield $A(v)$ unless a physical error has appeared there). The relevant measurements give us
\[\delta_s(c-w-k);\quad \delta_r(k+i+d-c-j-x+w-l);\quad \delta_t(-d+l+x)\]
for each new face, where $r,s,t\in \Z_d$. By substitution this gives
\[\delta_r(k+i+d-c-j-x+w-l) = \delta_r(-t-s+i-j) = \delta_{r+t+s}(i-j) = \delta_n(i-j) = \delta_i(n+j)\]
where $n$ is the group product of $r,t,s$ in $\Z_d$. Computationally, $n$ is the important \textit{measurement outcome} of the merge. Plugging back in to the patches we have
\[\input{merge_outcome_patch.tikz}\]
In the positive outcome case, i.e. when $s=r=t=0$, it is immediate that we have $|\delta_{g+h}\>_L$ on the combined patch. Otherwise, we can `fix' the internal additions of $s, t, n$ to the edges with string operators or alternatively accommodate them into the Pauli frame in the same manner as described in e.g. \cite{BH}. Then we are left with $q^{ng}|\delta_{g+h}\>_L$, as stated.

The Fourier transformed version of the above explains the rough merges as well, so we do not describe it explicitly.

\section{Proof of lattice counits}\label{app:counit}
We now show a `smooth counit' on a patch with state $|\delta_j\>_L$:
\[\input{delta_j_patch.tikz}\]
Measure out all edges in the $Z$ basis, giving
\[\sum_{a,b,c,d,i}q^{ij}\delta_r(a)\delta_s(a-c)\delta_t(c)\delta_u(i+b-a)\delta_v(b-d)\delta_w(i+d-c)\delta_x(-b)\delta_y(-d)\]
for some $r,\cdots,y\in \Z_d$. Then we observe that $\delta_u(i+b-a)=\delta_i(a-b-u)=\delta_i(n)$ for $n=a-b-u$, and by performing some other substitutions we arrive at
\[q^{nj}\delta_v(y-x)\delta_w(n-y-t)\delta_s(n-u-x-t)\]
Importantly, the only factor here which depends on the input state is $q^{nj}$. All the $\delta$-functions are merely conditions regarding which measurement outcomes are possible due to the lattice geometry. These will always be satisfied by our measurements, thus we have just
\[|\delta_j\>_L\mapsto q^{nj}\]
for $n\in\Z_d$, which in the other basis is $|i\>_L\mapsto \delta_{n,i}$ as stated. The rough counit follows similarly.

\section{Qudit ZX-calculus axioms}\label{app:zx_axioms}
We show some relevant axioms for the fragment of qudit ZX-calculus which interests us. These simply coincide with the rules from Hopf and Frobenius structures, along with the Fourier transform. We ignore the more general phase group \cite{W1}, and also leave out non-zero scalars. First, we define a spider
\[\input{spider_theorem.tikz}\]
which is well-defined due to associativity and specialty of the underlying Frobenius structure. The spider is also invariant under exchange of input wires with each other and the same for outputs, as the Frobenius algebra is (co)-commutative. A phaseless spider with 1 input and 1 output is identity:
\[\input{phaseless_spider.tikz}\]
Now, we can define duality morphisms on the object $\C\Z_d$, which we call a `cup' and similarly a `cap':
\[\input{cup.tikz}\]
which correspond to:
\[\input{cup_rules.tikz}\]
for the cup, and the vertically flipped version for the cap. The antipodes included here are responsible for the antipodes in the $CX$ gate in Section~\ref{sec:synth}. Then we have the Fourier exchange rule:
\[\input{fourier_exchange.tikz}\]
which encodes Lemma~\ref{lem:fourier} graphically.

Then we have the bialgebra rules
\[\input{bialgebra_rules.tikz}\]
and rules pertaining to the antipode:
\[\input{antipode_axioms.tikz}\]
This is far from an exhaustive set of rules.

\section{The logical block depiction}\label{app:block}
The lattice at a given time is drawn with a red line for a smooth boundary and green for a rough boundary:
\[\includegraphics[width=0.35\textwidth]{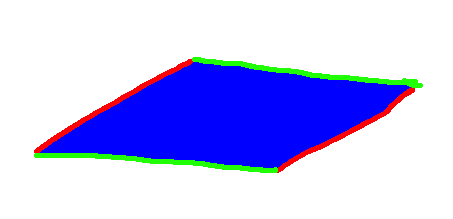}\]
where the surface is shaded blue for clarity. A block extending upwards represents the transformation over time. For example:
\[\includegraphics[width=0.1\textwidth]{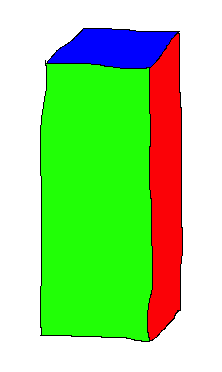}\]
We call this the `logical block' depiction, following similar work in \cite{Logic}.

Table~\ref{tbl:lat_oper} is an explicit dictionary between lattice surgery operations, qudit ZX-calculus and linear maps in the multiplicative fragment, i.e. the $n=0$ measurement outcomes. We choose to use the multiplicative fragment to highlight the visual connection between the columns. We see that red and green spiders correspond to rough and smooth operations respectively.

\begin{table}
  \centering
  \begin{tabular}{ | m{3cm} | c | m{2cm} | m{3cm} | }
    \hline
    Lattice operation & Logical block & ZX-diagram & Linear map\\ \hline
	smooth unit
	&
    \begin{minipage}{.05\textwidth}
      \includegraphics[width=\linewidth]{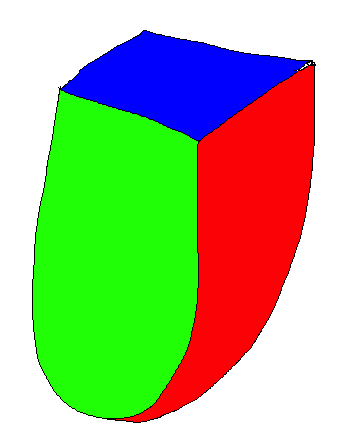}
    \end{minipage}
    &
      \[\input{smooth_unit_ZX.tikz}\]
    & 
    \[\sum_i|i\>\]
	\\
	\hline
	smooth split
	&
	\begin{minipage}{.1\textwidth}
      \includegraphics[width=\linewidth]{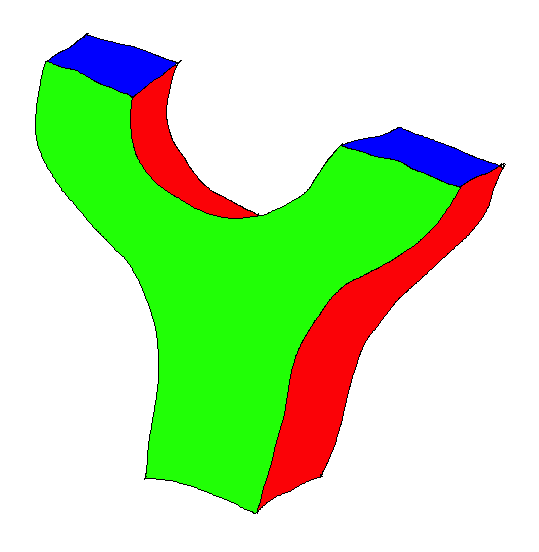}
    \end{minipage}
    &
     \[\input{smooth_split_ZX.tikz}\]
    &
	\[|i\>\mapsto |i\>\tens|i\>\] 
    \\ 
	\hline
	smooth merge
	&
	\begin{minipage}{.1\textwidth}
      \includegraphics[width=\linewidth]{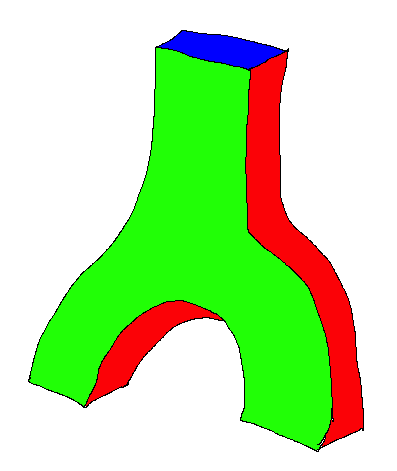}
    \end{minipage}
    &
      \[\input{smooth_merge_ZX.tikz}\]
    & 
    \[|i\>\tens|j\>\mapsto \delta_{i,j}|i\>\]
	\\
	\hline
	smooth counit
	&
	\begin{minipage}{.05\textwidth}
      \includegraphics[width=\linewidth]{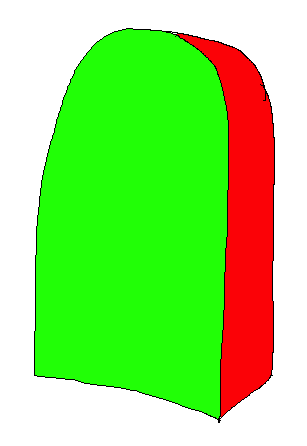}
    \end{minipage}
    &
      \[\input{smooth_counit_ZX.tikz}\]
    & 
    \[|i\>\mapsto 1\]
	\\
	\hline
	rough unit
	&
	\begin{minipage}{.05\textwidth}
      \includegraphics[width=\linewidth]{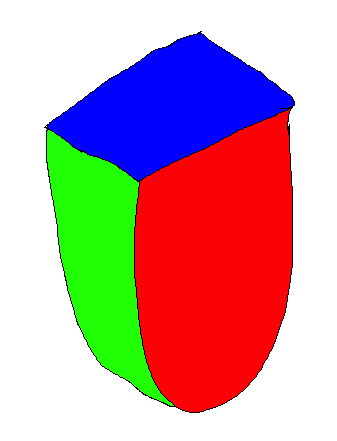}
    \end{minipage}
    &
      \[\input{rough_unit_ZX.tikz}\]
    & 
    \[|0\>\]
	\\
	\hline
	rough split
	&
	\begin{minipage}{.1\textwidth}
      \includegraphics[width=\linewidth]{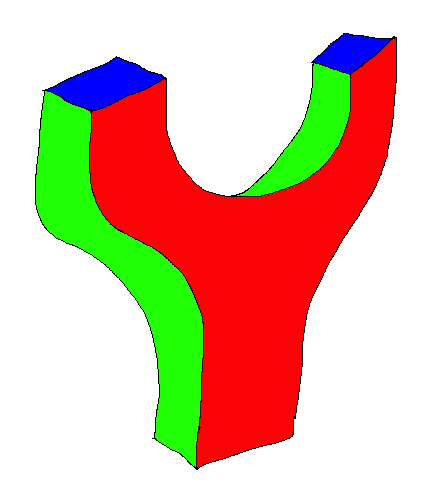}
    \end{minipage}
    &
      \[\input{rough_split_ZX.tikz}\]
    & 
    \[|i\>\mapsto \sum_h|h\>\otimes |i-h\>\]
	\\
	\hline
	rough merge
	&
	\begin{minipage}{.1\textwidth}
      \includegraphics[width=\linewidth]{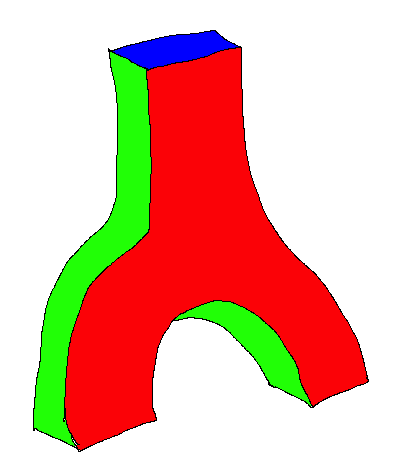}
    \end{minipage}
    &
      \[\input{rough_merge_ZX.tikz}\]
    & 
    \[|i\>\tens|j\>\mapsto |i+j\>\]
	\\
	\hline
	rough counit
	&
	\begin{minipage}{.05\textwidth}
      \includegraphics[width=\linewidth]{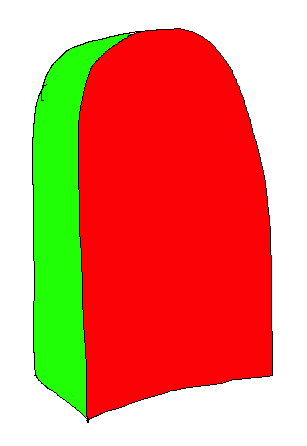}
    \end{minipage}
    &
      \[\input{rough_counit_ZX.tikz}\]
    & 
    \[|i\>\mapsto\delta_{i,0}\]
	\\
	\hline
	rotation
	&
	\begin{minipage}{.05\textwidth}
      \includegraphics[width=\linewidth]{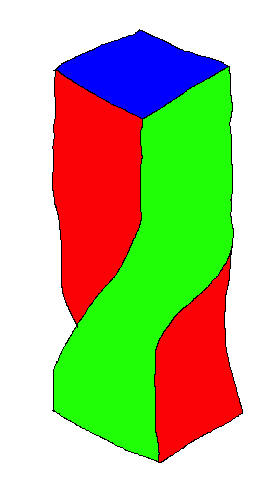}
    \end{minipage}
    &
      \[\input{fourier_spider.tikz}\]
    & 
    \[|i\>\mapsto\sum_jq^{-ij}|j\>\]
	\\
	\hline
  \end{tabular}
  \caption{Dictionary of lattice surgery operations in the multiplicative fragment.}\label{tbl:lat_oper}
\end{table}

We have no new results or proofs in this section, but we would like to discuss the diagrams of logical blocks. These sorts of diagrams for lattice surgery have been used in an engineering setting to compile quantum circuits to lattice surgery \cite{GF,Logic}. To go from the cubes shown there to the tubes which we show here we merely relax the discretisation of space and time somewhat to expose the relationship with algebra. This relationship with algebra is relevant because such diagrams have appeared in a seemingly quite different context. 

It is well known that the category of `2-dimensional thick tangles', \textbf{2Thick}, is monoidally equivalent to the category \textbf{Frob} freely generated by a noncommutative Frobenius algebra \cite{Lauda}. This should be unsurprising to those familiar with the notion of a `pair of pants' algebra. We say that \textbf{2Thick} is a \textit{presentation} of \textbf{Frob}. Similarly, the symmetric monoidal category \textbf{2Cob} of (diffeomorphism classes of) 2-dimensional cobordisms between (disjoint unions of) circles is a presentation of \textbf{ComFrob}, the category freely generated by a commutative Frobenius algebra \cite{Kock}.

This fact is important for topological quantum field theories (TQFTs). One can define an $n$-dimensional TQFT as a symmetric monoidal functor from $\textbf{nCob}\rightarrow \textbf{Vect}$, the category of finite-dimensional vector spaces. The key point is that the functor takes (diffeomorphism classes of) manifolds as inputs and outputs linear maps between vector spaces, which are by definition manifold invariants. One can see that 2D TQFTs are in bijection with commutative Frobenius algebras in $\textbf{Vect}$.

In \cite{Reut}, Reutter gives a slightly different monoidal category, which we will call \textbf{2Block}. It has as objects disjoint unions of squares, with the same shading of sides as those in the logical block diagrams above. Then morphisms are classes of surfaces between the squares, such that the borders between the surfaces match up with the edges of the squares at the source and target objects and the surface colours are consistent with those of the squares' sides. While the morphisms are obviously quotiented by equivalence of surfaces up to border-preserving diffeomorphism, Reutter quotients by `saddle-invertibility' as well, which is not a rule one can acquire through topological moves alone, as it involves the closing and opening of holes.

Reutter conjectures that $\textbf{2Block}\simeq \textbf{uHopf}$, where \textbf{uHopf} is the category freely generated by a unimodular Hopf algebra.\footnote{In \textbf{Vect}, unimodularity is typically defined using integrals \cite{Ma:book}. In this more abstract setting it is defined by some axioms on dualities.} While we do not know enough about topology or geometry to prove (or disprove) this conjecture, we suspect one route is to consider Morse functions and classify the diffeomorphism classes near critical points. This is similar to one proof of $\textbf{2Cob}\simeq \textbf{ComFrob}$ \cite{Kock}. For the reader's convenience, we now reproduce a handful of the equivalences under topological deformation which motivate this conjecture. We have the axioms of a Frobenius algebra,
\[\includegraphics[width=0.3\linewidth]{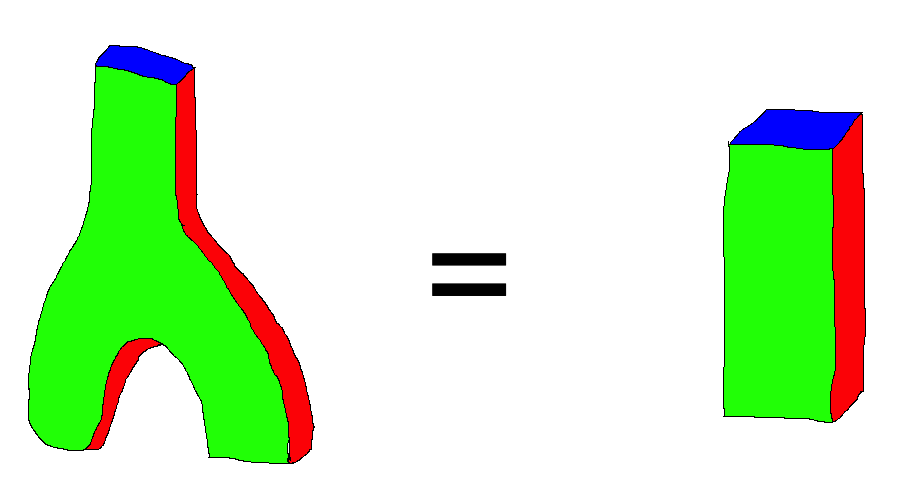}\]
\[\includegraphics[width=0.3\linewidth]{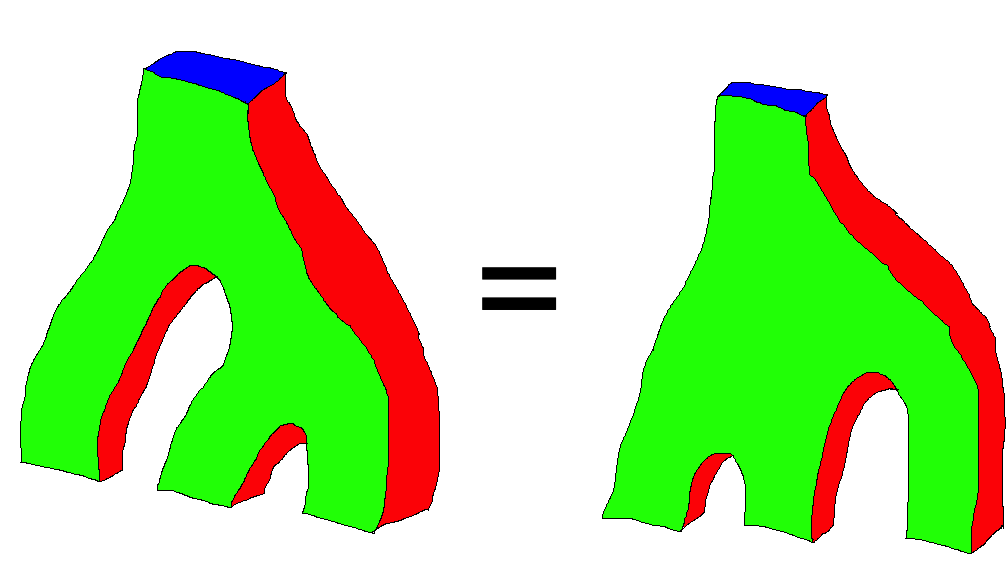}\]
\[\includegraphics[width=0.3\linewidth]{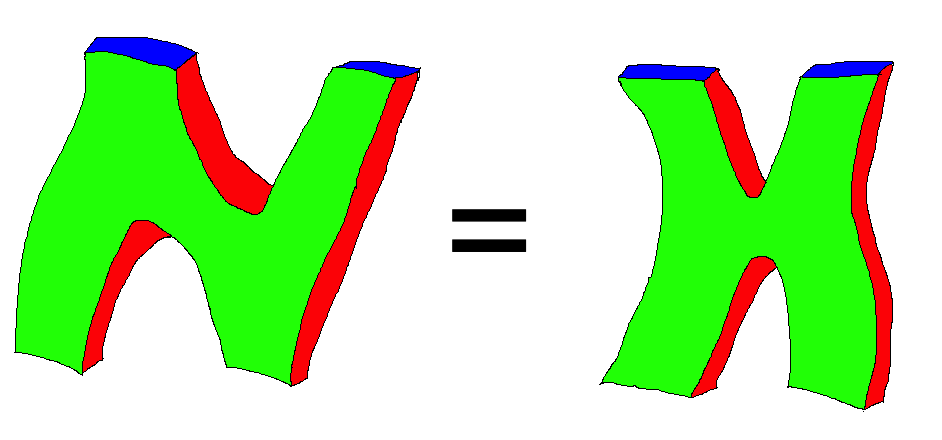}\]
and the same for red faces. These are just widened versions of the diagrams in \textbf{2Thick}. Then one can see the interpretation of a unimodular Hopf algebra as two interacting Frobenius algebras. We start with two Frobenius algebras and glue them together in such a way that they give the bialgebra and antipode axioms. The main bialgebra rule is
\[\includegraphics[width=0.3\linewidth]{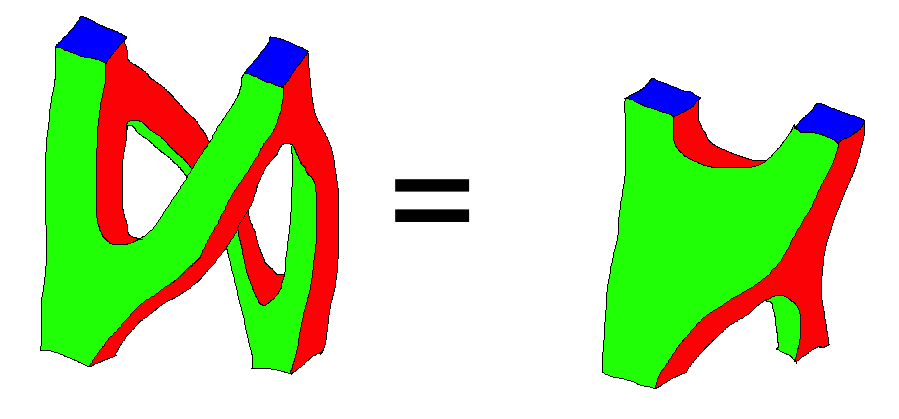}\]
where we require saddle invertibility to close up a hole in the middle. This is also required for showing that comultiplication is a unit map and so on. Given all of these deformations and those involving the antipode, which is a twist by $\pi$, one can see that they define a functor $\textbf{uHopf}\rightarrow\textbf{2Block}$; the hard part is proving that this is an equivalence.

Now, Reutter also draws a comparison with representation theory and tensor category theory. It is striking that, given the unimodular Hopf algebra $\C\Z_d$, we can create a logical space on a patch isomorphic to the vector space of $\C\Z_d$ itself, and the logical operations precisely coincide with the linear maps defined by the algebra. We conjecture that lattice surgery is the `computational implementation' of this presentation of unimodular Hopf algebras, in the same way that the logical space of the Kitaev model on a closed orientable manifold $\mathcal{M}$ is isomorphic to the vector space $F(\mathcal{M})$ in the image of a Dijkgraaf-Witten theory $F : \textbf{2Cob}\rightarrow \textbf{Vect}$ when given the same manifold $\mathcal{M}$ \cite[Thm~3.2]{Cow}. It remains to be seen whether this extends further than just abelian group algebras.

\section{Logical $CX$ gate}\label{app:cx}
Here we check the correctness of the $CX$ gate implementations from Section~\ref{sec:synth}.

First, observe that the diagram:
\[\input{left_cnot.tikz}\]
yields the linear map:
\[|i\>\otimes|j\>\mapsto |i\>\otimes |i\> \otimes |j\> \mapsto |i\>\otimes |i+j\>\]
where we have considered the diagram piecemeal from bottom to top, indicated by the dashed lines.

Then we can perform a sequence of rewrites between all four diagrams, labelled below:
\[\input{rewrite_cnot1.tikz}\]
where at each stage we have either used the spider rule, inserted duals, or swapped between duals and spiders; see Appendix~\ref{app:zx_axioms}.

\section{Generalisations and Hopf algebras}\label{app:generalisations}
While we have shown that lattice surgery works for arbitrary dimensional qudits, we emphasise that the algebraic structures involved are very simple so far. The lattice model in the bulk can be generalised significantly: first, one can replace $\C\Z_d$ with another finite abelian group algebra. As all finite abelian groups decompose into direct sums of cyclic groups this case follows immediately from the work herein and is uninteresting. 

At the second level up, we can replace it with an arbitrary finite group algebra $\C G$. At this level several assumptions break down: 
\begin{itemize}
\item $\C G$ still has a dual function algebra $\C(G)$, but the Fourier transform no longer coincides with Pontryagin duality, and the two algebras will no longer be isomorphic in general. One can still define a Fourier transform in the sense that it translates between convolution and multiplication, but in this case the Fourier transform is the Peter-Weyl isomorphism, i.e. a bimodule isomorphism between $\C G$ and a direct sum of matrix algebras labelled by the irreps of $G$.
\item The $\C G$ lattice model can no longer be described using string operators, and these must be promoted to ribbon operators \cite{Kit1}. This is because the lattice model is based on the Drinfeld double $D(G) = \C(G)\lcross \C G$, where the associated action is conjugation. In the abelian case conjugation acts trivially and so we have $D(\Z_d) = \C(\Z_d) \otimes \C\Z_d$: the double splits into independent algebras, which give the $X$-type and $Z$-type string operators respectively.
\item There are still canonical choices of rough and smooth boundary, labelled by subgroups $K = \{e\}$ and $K = G$ for rough and smooth boundaries respectively. Similarly, we still have well-defined measurements, using representations of $C G$ and $C(G)$ for vertices and faces. However, the algebra of ribbon operators which are undetectable at the boundary, and hence the logical operations on a patch, becomes significantly more complicated, see \cite{PS2} for the underlying module theory. Preliminary calculations indicate that they are labelled by conjugacy classes (i.e. irreps) of $G$, and it is not even obvious that ${\rm dim}(\CH_{vac}) = |G|$ as in the abelian case. This is quite an obstruction to calculating the logical maps corresponding to lattice surgery operations.
\end{itemize}

Of course, the Kitaev model can be generalised much further still. The third level would be arbitrary finite-dimensional Hopf $\C^*$-algebras. At this level even the calculations in the bulk are tricky, and many features were only recently resolved \cite{Cow,Meu,Chen}. Understanding lattice surgery in these models seems a formidable task. We aim to at least make some progress on this in upcoming work \cite{Cow2}. 

The fourth (and highest) level is the maximal generality, which are weak Hopf $\C^*$-algebras, in bijection (up to an equivalence) with so-called \textit{unitary fusion categories} \cite{EGNO}. Even at this extreme generality, there are glimpses of hope. There are two canonical choices of boundaries given by the trivial (rough) and regular (smooth) module categories \cite{Os}, and we speculate that calculating some basic features like ${\rm dim}(\CH_{vac})$ of a patch could be done using techniques from topological quantum field theory (TQFT). At this level of generality, the connections with TQFT become more tantalising. The parallels between topological quantum computing in the bulk and TQFTs are well-known, see e.g. \cite{Kir}, but lattice surgery introduces discontinuous deformations in the manner of geometric surgery. While boundaries of TQFTs are well-studied \cite{KS,FS}, we do not know whether TQFT theorists study the relation between geometric surgery on manifolds and linear algebra in the same manner as they do for, say, diffeomorphism classes of cobordisms.

\bibliographystyle{eptcs}
\end{document}